\documentclass{desyproc}

\begin{document}

\title{Searches for axioelectric effect of solar axions with BGO-scintillator and BGO-bolometer detectors}

\author{{\slshape V.N. Muratova$^1$, A.V. Derbin$^1$, L. Gironi$^{2,3}$, S.S Nagorny$^{4,5}$, L. Pattavina$^4$, S.V. Bakhlanov$^1$,
 J.W. Beeman$^6$, F.Bellini$^{7,8}$, M. Biassoni$^{2,3}$, S. Capelli$^{2,3}$, M. Clemenza$^{2,3}$, I.S. Dratchnev$^{1,5}$,
  E. Ferri$^{2,3}$, A. Giachero$^{2,3}$, C. Gotti$^{2,3}$, A.S. Kayunov$^1$, C. Maiano$^{2,3}$, M. Maino$^{2,3}$,
  M. Pavan$^{2,3}$, S. Pirro$^4$, D.A. Semenov$^1$, M. Sisti$^{2,3}$, E.V. Unzhakov$^1$}\\ [1ex]
$^1$St.Petersburg Nuclear Physics Institute, Gatchina 188350 - Russia\\
$^2$INFN - Sezione di Milano Bicocca, Milano 1-20126 - Italy\\
$^3$Dipartimento di Fisica, Universita di Milano-Bicocca, Milano 1-20126 - Italy\\
$^4$INFN - Laboratory Nazionale del Gran Sasso, Assergi (L'Aquila) I-67100 - Italy\\
$^5$Gran Sasso Science Institute, INFN, L'Aquila (AQ) I-67100 - Italy\\
$^6$Lawrence Berkeley National Laboratory, Berkley, California 94720 - USA\\
$^7$INFN -Sezione di Roma, Roma I-00185 - Italy\\
$^8$Dipartimento di Fisica - Universita di  Roma La Sapienza, Roma I-00185 - Italy}

\contribID{muratova\_valentina}

\confID{300768} 
\desyproc{DESY-PROC-2014-03}
\acronym{Patras 2014} 
\doi 
\maketitle

\begin{abstract}
A search for axioelectric absorption of 5.5 MeV solar axions produced in the $p + d \rightarrow {^3\rm{He}}+\gamma~(5.5~
\rm{MeV})$ reaction has been performed with a BGO detectors. A model-independent limit on the product of axion-nucleon $g_{AN}^3$
and axion-electron $g_{e}$ coupling constants has been obtained: $| g_{Ae}\times g_{AN}^3|< 1.9\times 10^{-10}$ for 90\% C.L..
\end{abstract}
\section{Introduction}

There are new possibilities for strong CP problem solution, which allow the existence of axions with a large mass (~1 MeV), while
their interaction with ordinary particles remain at the level of the invisible axions. The models rely on the hypothesis of
mirror particles \cite{Ber01} and SUSY at the TeV scale \cite{Hal04}. The existence of these heavy axions is not forbidden by the
laboratory experiments or astrophysical data.

 This article describes the experimental search for 5.5 MeV solar axions, which can  be produced by $p + d
\rightarrow\rm{^3He}+ A$ reaction. Axion flux should be proportional to the $pp$-neutrino flux, which has been estimated with
high accuracy \cite{Bel14}. The searches have been performed with the use of bismuth orthogermanate $\rm{Bi_4Ge_3O_{12}}$ (BGO)
scintillation and bolometric detectors.  The solar axions are supposed to interact with atoms via the reaction of axioelectric
effect ${\rm A}+e+Z\rightarrow e+Z$. This kind of interaction is governed by $g_{Ae}$-constant and the cross section depends on
the charge as $Z^5$. From this point of view the BGO detector is a very suitable target, because of the high $Z_{Bi} = 83$ of
bismuth nucleus.

The high energy solar axions and axions from a nuclear reactor have been looked by the Borexino \cite{Bel08,Bel12}, the CAST
\cite{And10} and the Texono \cite{Cha07} collaborations. This paper  is based on the results of obtained with BGO scintillation
\cite{Der13} and BGO bolometer detectors \cite{Der14}.

\begin{figure}
\centerline{\includegraphics[bb = 20 120 500 755, width=0.45\textwidth, height=0.3\textheight]{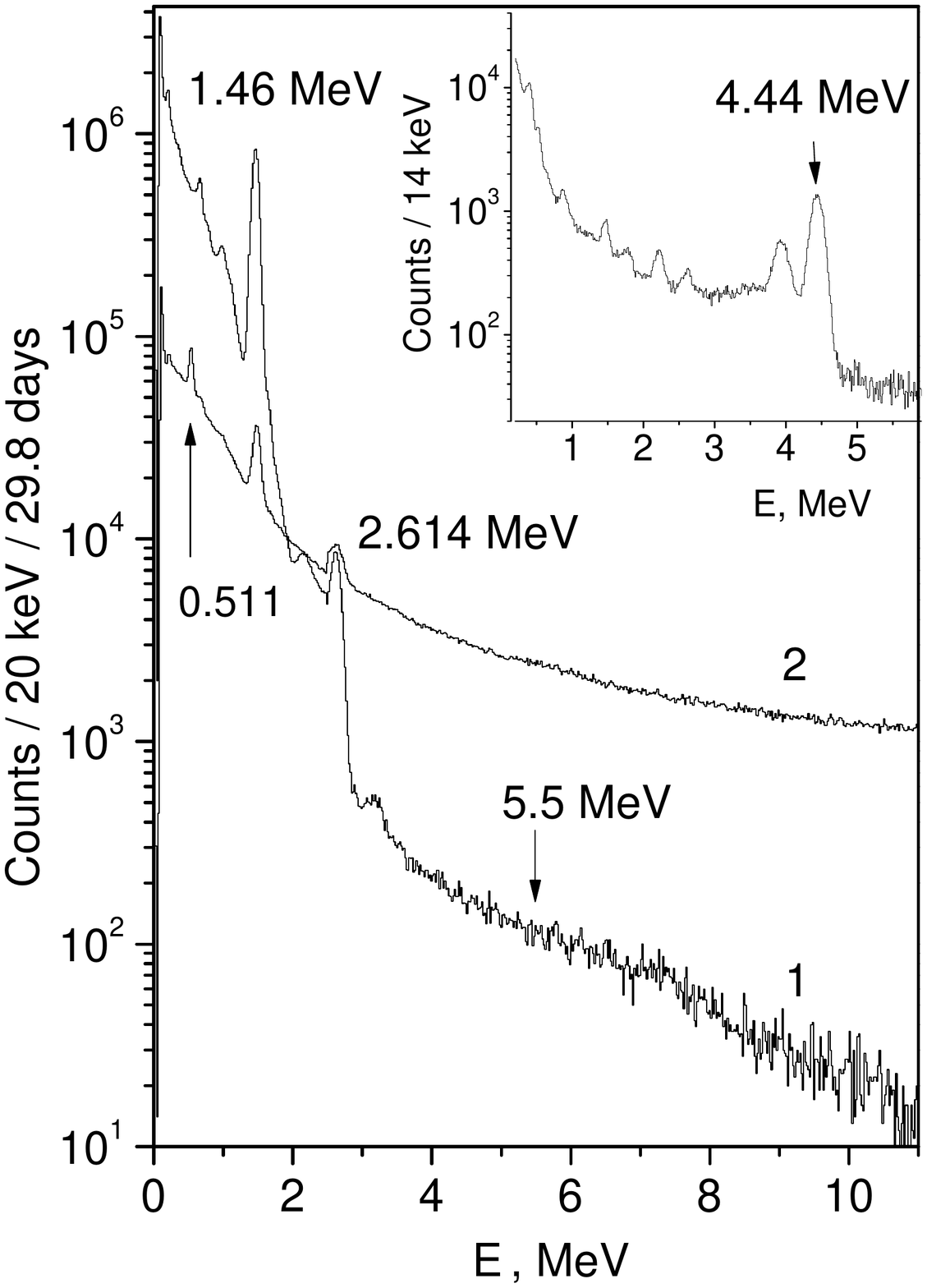}}
\caption {\small{The energy spectrum of the BGO detector measured (1) in anticoincidence and (2) in coincidence with the muon
veto signal.}} \label{Muratova_Valentina_fig1.pdf}
\end{figure}
The 5.5 MeV axion production probability ratio $(\omega_A / \omega_\gamma)$ depends only on the isovector axion-nucleon coupling
constant $g^3_{AN}$ (see \cite{Der14} and refs therein):
\begin{equation}\label{ratio}
\frac{\omega_{A}}{\omega_{\gamma}} =
 \frac{\chi}{2\pi\alpha}\left[\frac{g_{AN}^{3}}{\mu_3}\right]^2\left(\frac{p_A}{p_\gamma}\right)^3 = 0.54(g_{AN}^{3})^2
 \left(\frac{p_A}{p_\gamma}\right)^3.
\end{equation}
where $p_{\gamma}$ and $p_{A}$ are, respectively, the photon and axion momenta and $\mu_3$ is isovector nuclear magnetic momenta.
 At the Earth's surface the axion flux is:
\begin{eqnarray}\label{FluxA}
\Phi_A = \Phi_{\nu p p}(\omega_A/\omega_\gamma)
\end{eqnarray}
where $\Phi_{\nu p p}$ is the $pp$-neutrino flux. The cross section for the a.e. effect was calculated in \cite{Zhi79}.

\section{BGO scintillation detector}
BGO crystal with mass 2.46 kg contains 1.65 kg of Bi. The crystal was shaped as a cylinder, 76 mm in diameter and 76 mm in
height. The detector signal was measured by an R1307 photoelectron multiplier, which had an optical contact with a crystal end
surface. The external $\gamma$ activity was suppressed using a passive shield that consisted of layers of lead and bismuth
(${\rm{Bi}}_2{\rm{O}}_3$) with the total thickness $\approx$ 110 ${\rm{g\; cm}}^{-2}$. The setup was located on the Earth's
surface. In order to suppress the cosmic-ray background we used an active veto, which consisted of five $50\times50\times12$ cm
plastic scintillators.  The measurements were performed over 29.8 days. The energy spectrum of the BGO detector in the range of
(0--11) MeV is shown in Fig. 1. In inset the calibration spectrum measured with Pu-Be neutron source is shown.

In the spectrum, one can identify two pronounced peaks at 1.460 MeV and 2.614 MeV; these are due to the natural radioactivity of
the $^{40}{\rm{K}}$ and of $^{208}{\rm{Tl}}$ from the $^{232}{\rm{Th}}$ family.  The positions and intensities of these peaks
were used for monitoring of time stability of the detector.

\section{BGO bolometers}
Four cubic ($5\times5\times5$ $\rm{cm}^3$) BGO bolometers, containing 1.65 kg of Bi, were arranged in a four-plex module, one
single plane set-up. The scintillation light was monitored with an auxiliary bolometer made of high-purity germanium
\cite{Bee13}. The detector was installed in the $^3\rm{He}/^4\rm{He}$ dilution refrigerator in the underground laboratory of
L.N.G.S. and operated at a temperature of few mK. The four crystals were housed in a highly pure copper structure, the same
described in \cite{Ale12}. Neutron Transmutation Doped (NTD) germanium thermistor was coupled to each bolometer, NTD acts as a
thermometer: recording the temperature rises produced by particle interaction and producing voltage pulses proportional to the
energy deposition. Details on electronics and on the cryogenic set-up can be found in \cite{Pir00,Arn06}. The detector was
operated for a total live time of 151.7 days.


\begin{figure}
\centerline{\includegraphics[bb = 20 120 500 755, width=0.45\textwidth, height=0.3\textheight]{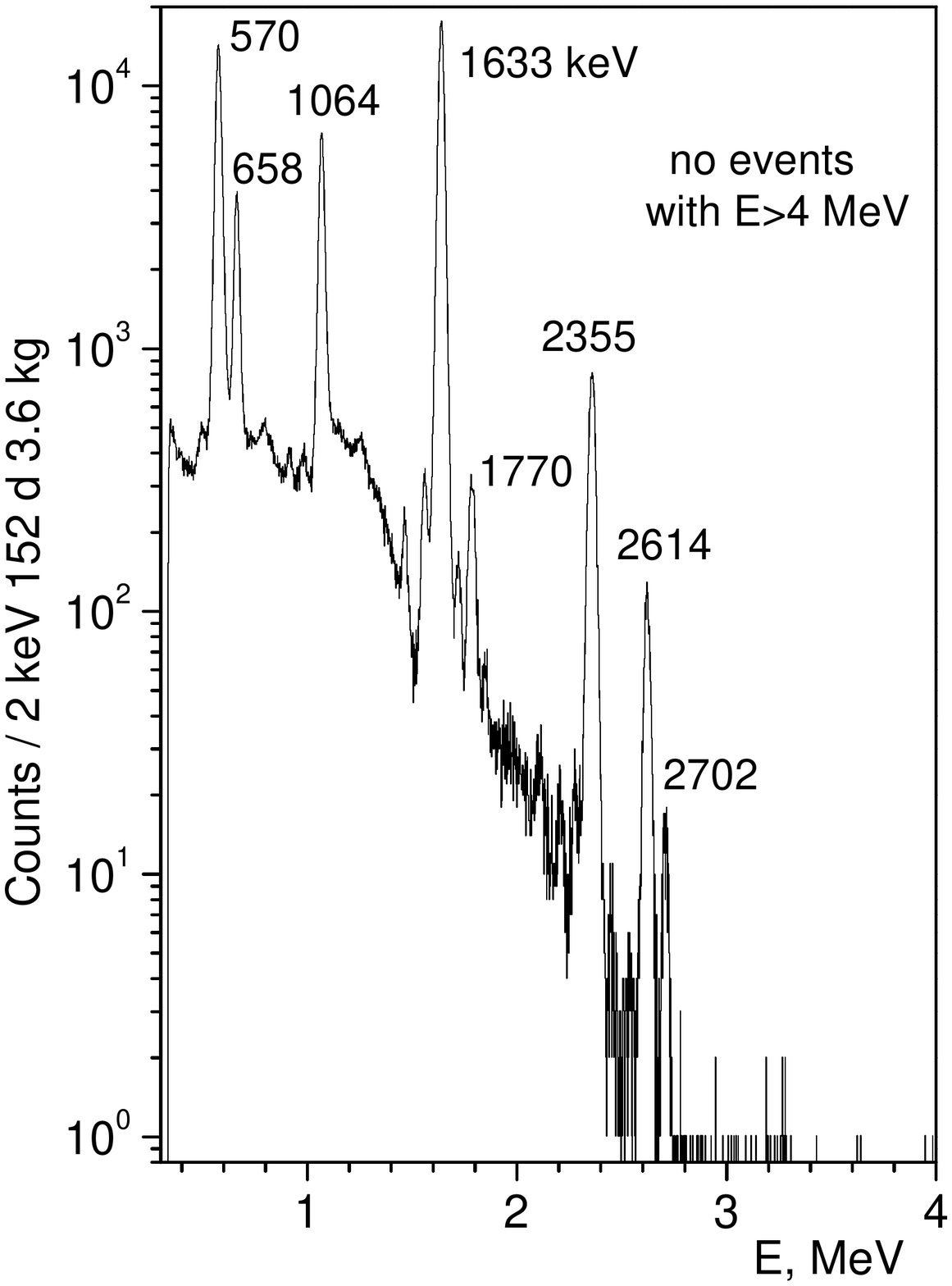}}
\caption {The energy spectrum of the four BGO detectors ($\beta- \rm{and}~\gamma-$ events) measured for a 151.7 days. }
\label{Muratova_Valentina_fig2.pdf}
\end{figure}


The different light yield of interacting particles was used to discriminate $\alpha-$events from $\beta/\gamma$ ones. This
allowed to strongly increase the sensitivity, due to rejection of all $\alpha-$events in the region of interest. The total
statistics in the range of (0.3-4) MeV for $\beta/\gamma$ events are in shown in Fig.\ref{Muratova_Valentina_fig2.pdf}. The most
intense gamma lines are produced by $^{207}\rm{Bi}$ decays. In first approximation, the energy resolution of large mass
bolometric detector is independent of the energy - the FWHM is $33.7\pm0.6$ keV at 2614 keV ($^{208}\rm{Tl}$) and $33.2\pm0.1$
keV at 570 keV ($^{207}\rm{Bi}$).

\section{Results}

The expected number of axioelectric absorption events are:
\begin{equation}
S_{abs} = \varepsilon N_{Bi}T\Phi_A\sigma_{Ae}
\end{equation}
where $\sigma_{Ae}$ is the axioelectric effect cross section; $\Phi_A$ is the axion flux (\ref{FluxA}); $N_{Bi}$  is the number
of Bi atoms; $T$  is the measurement time; and $\varepsilon$ is the detection efficiency for 5.5 MeV electrons.

\begin{figure}
\centerline{\includegraphics[bb = 20 120 500 755, width=0.45\textwidth, height=0.3\textheight]{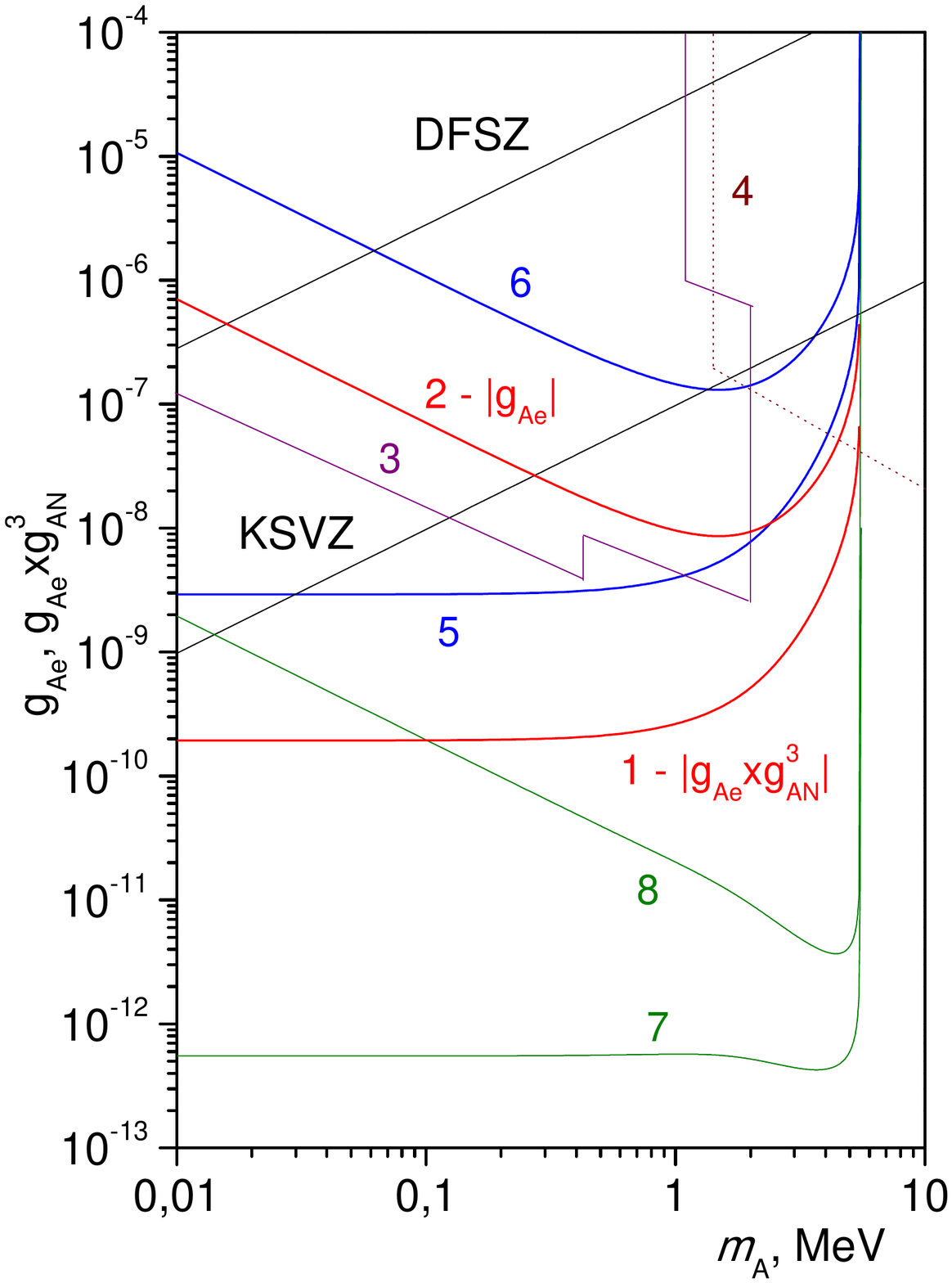}}
\caption{The limits on the $g_{Ae}$ coupling constant obtained by 1,2 -BGO-bolometer \cite{Der14} and 5,6-BGO-scintillator
\cite{Der13} for $|g_{Ae}|$ and $|g_{Ae}\times g^3_{AN}|$, correspondingly; 3- solar and reactor experiments, 4- beam dump
experiments; 7,8 - Borexino results for $|g_{Ae}\times g^3_{AN}|$ and $|g_{Ae}|$ \cite{Bel12}.}
\label{Muratova_Valentina_fig3.pdf}
\end{figure}
The axion flux $\Phi_A$ is proportional to the constant $(g^3_{AN})^2$, and the cross section $\sigma_{Ae}$ is proportional to
the constant $g^2_{Ae}$. As a result, the $S_{abs}$ value depends on the product $(g_{Ae})^2\times (g^3_{AN})^2$. The
experimentally found condition $S_{abs} \leq S_{lim}$ imposes some constraints on the range of  possible $|g_{Ae}\times
g^3_{AN}|$  and $m_A$ values.

The range of excluded $|g_{Ae}\times g^3_{AN}|$ values is shown in Fig.~3, at $m_A \rightarrow 0$ the limits
\begin{equation}
|g_{Ae}\times g^3_{AN}| \leq 2.9\times10^{-9}~~\rm{and}
\end{equation}
\begin{equation}
|g_{Ae}\times g^3_{AN}| \leq 1.9\times10^{-10}~~\rm{at~90\%~c.l.}.\label{limgaegan}
\end{equation}
were obtained for BGO scintillating and BGO bolometer detectors, correspondingly. These constraints are completely
model-independent and valid for any pseudoscalar particle with coupling $|g_{Ae}|$ less than $10^{-6(4)}$.

For hadronic axion model with concrete relation between $g_{AN}^3$ and $m_A$ one can obtain a constraint on the $g_{Ae}$
constant, depending on the axion mass (Fig.~3). For $m_A$ = 1 MeV, this limit corresponds to $|g_{Ae}|\leq 9.6 \times 10^{-9}$ at
$\rm{90\% c.l.}$.

Figure 3 also shows the constraints which were obtained in the Borexino experiment for 478 keV ${^7\rm{Li}}$ solar axions
\cite{Bel08} and in the Texono reactor experiment for 2.2 MeV axions produced in the $n + p \rightarrow d + A$ reaction
\cite{Cha07}. Borexino coll. reported new more stringent limits on $g_{Ae}$ coupling for 5.5 MeV solar axions \cite{Bel12}.
Unlike our work, these limits on $g_{Ae}$ coupling were obtained in assumption that the axion interacts with electron through the
Compton conversion process.

In the model of the mirror axion \cite{Ber01} an allowed parameter window is found within the P-Q scale $f_A\sim10^4-10^5$ GeV
and the axion mass $m_A\sim$ 1 MeV. The limit (\ref{limgaegan}) may be represented as a limit on the value $f_A$ by taking the
following relations into account: $g_{AN}^3=0.5(g_{Ap}-g_{An})= 1.1/f_A$ and $g_{Ae}=5\times10^{-4}/f_A$. For axion masses about
1 MeV, the limit is $f_A > 1.7\times10^{3} \rm{~GeV}$, which is close to the lower bound of  mirror axion window.

Our results set constraints on the parameter space of the CP-odd Higgs $(A^0)$, which arise in the next-to-minimal supersymmetric
Standard Model due to the spontaneous breaking of approximate symmetries such as PQ-symmetry. The corresponding exclusion region
can be obtained from Fig.\ref{Muratova_Valentina_fig3.pdf} using the conversion $C_{Aff} = g_{Ae}2m_W/g_2 m_e$ where $C_{Aff}$ is
the coupling of the CP-odd Higgs to fermions and $g_2=0.62$ is the gauge coupling. The limit (\ref{limgaegan}) translates into
$C_{Aff}\times m_{A^0} \leq 3\times 10^{-3}$ MeV for $m_{A^0} < 1$ MeV, which is compatible with the limits obtained in reactor
experiments exploring Compton conversion.

This work was supported by RFBR grants 13-02-01199 and 13-02-12140-ofi-m.


\begin{thebibliography}{21}


\bibitem{Ber01} Z.~Berezhiani et al., Phys.~ Lett. B500, 286 (2001).
\bibitem{Hal04} L.J.~Hall and T.~Watari, Phys.~Rev. D70, 115001 (2004).
\bibitem{Bel14} G.~Bellini et al., (Borexino coll.) Nature, 512, 7515, 383 (2014)
\bibitem{Bel08} G.~Bellini et al., (Borexino coll.) EPJ, C54, 61 (2008).
\bibitem{Bel12} G.~Bellini et al., (Borexino Coll.), Phys. Rev. D 85, 092003 (2012).
\bibitem{And10} S.~Andriamonje et al., (CAST coll.) JCAP 1003, 032, (2010). arXiv:0904.2103
\bibitem{Cha07} H.M.~Chang et al., (Texono Coll.) Phys.~ Rev. D75, 052004 (2007).
\bibitem{Der13} A.V.~Derbin et al., Europ.~Phys.~J. C73, 2490 (2013). arXiv:1306.4574
\bibitem{Der14} A.V.~Derbin et al., Europ.~Phys.~J. C74, 3035 (2014). arXiv:1405.3782
\bibitem{Zhi79} A.R.~Zhitnitskii and Yu.I.~Skovpen, Yad. Fiz., 29b, 995 (1979).
\bibitem{Car12} L.~Cardani, S.Di~Domizio, L.~Gironi, JINST 7, P10022 (2012).
\bibitem{Bee13} J.W.~Beeman et al., JINST 8, P05021 (2013).
\bibitem{Ale12} F.~Alessandria et al., Astropart. Phys. 35, 839�849 (2012).
\bibitem{Pir00} S.~Pirro et al., Nucl. Instrum. Methods A 444, 331 (2000).
\bibitem{Arn06} C.~Arnaboldi et al., Nucl. Instrum. Methods A 559, 826 (2006).

\end{thebibliography}
\end{document}